\documentclass[prl,twocolumn,showpacs,floatfix,amsmath,amssymb,superscriptaddress]{revtex4-2}
\usepackage{amsfonts}
\usepackage{stmaryrd}
\usepackage{bbm}
\usepackage{mathrsfs}
\usepackage{tipa}
\usepackage{amssymb}
\usepackage{txfonts}
\usepackage{graphicx}
\usepackage{dcolumn}
\usepackage{epstopdf}
\usepackage[colorlinks,linkcolor=blue,urlcolor=black,citecolor=blue]{hyperref}
\usepackage{multirow}
\usepackage{subfigure}
\usepackage{url}
\usepackage{upgreek}
\usepackage[utf8]{inputenc}
\usepackage[english]{babel}

\begin{document}

\newcommand*{\cm}{cm$^{-1}$\,}
\newcommand*{\Tc}{T$_c$\,}




\title{Distinct terahertz third-harmonic generation of many-body excitonic states}

\author{Changqing Zhu}
\author{Anneke Reinold}
\author{Patrick Pilch}
\author{Sergey Kovalev}
\author{Julian~Heckötter}
\author{Doris Reiter}
\author{Manfred Bayer}
\author{Marc Assmann}
\author{Zhe Wang}
\email{zhe.wang@tu-dortmund.de}
\affiliation{Department of Physics, TU Dortmund University, 44227 Dortmund, Germany}


\begin{abstract}
The dynamics of an electron-hole plasma governed by strong Coulomb interaction is a challenging many-body problem.
We report on experimental realization of electron-hole many-body states in the picosecond time scale, with tunable densities in a representative semiconductor Cu$_2$O. 
By using time-resolved optical-pump terahertz third-harmonic-generation spectroscopy, we study the nonlinear terahertz dynamical characteristics of the many-body electron-hole states.  
We find not only efficient and nonperturbative terahertz third-harmonic yield associated with the excitonic formation, but also a nonmonotonic dependence of the excitonic nonlinear response on the electron-hole density, reflecting the exciton dissociation at high charge density.
Our results provide an efficient excitonic sensing of the far-from-equilibrium electron-hole many-body states.
\end{abstract}

\maketitle


Excitonic dynamics plays a key role in various contexts, such as photosynthesis in chromophore molecules \cite{Fassioli14}, energy transport in semiconductor optoelectronic devices \cite{Mueller2018}, and development of novel quantum technologies \cite{Wilson2021}.
Similar to a hydrogen atom stabilized by attractive Coulomb interaction of an electron and a proton, an exciton is a quasiparticle formed as an attractively bound state of a negatively charged electron and a positively charged hole.
To observe the hydrogenic Rydberg series of an exciton is, however, not straightforward, because in a real material a concomitant excitation of unbound electron-hole pairs is hardly avoidable and the band dispersion often deviates from parabolic  \cite{Nature2014giant,PRB2021giant,Chernikov14}.
While the hydrogenic single-body description remains valid in the dilute charge density limit \cite{Heckoetter18}, the existence of exciton-exciton and exciton-plasma interactions renders the excitonic dynamics a representative many-body problem \cite{KaltKlingshirn24,Huber2001,Kaindl2003}.

From the time-dependent kinetic perspective, many-body effects do not emerge instantaneously after a femtosecond laser pulse excitation (1~fs = $10^{-15}$~s) of unbound electron-hole pairs, but builds up in a characteristic time scale of several 10~fs up to 0.1~picosecond (1~ps = $10^{-12}$~s) \cite{Huber2001}.
The many-body interactions govern the follow-up complex dynamical processes, before the system relaxes to a quasi-equilibrium state that consists of mainly the lowest-energy $1s$ excitons, which takes typically several 100~ps \cite{Kaindl2003}.
While the quasi-equilibrium state can be well approximated by an independent quasiparticle picture, it is a very challenging problem to describe the strongly Coulomb correlated many-body dynamics from 0.1 to 10~ps after the initial excitation \cite{Kira98}.
In this ultrashort time interval, various dynamical processes occur simultaneously, including the formation of excitonic states from unbound electron-hole pairs, the Coulomb collisions of one exciton with another bound or unbound electron-hole pair, as well as scattering of excitons or unbound charges by collective excitations (for example, phonons or plasmons), see e.g. \cite{KaltKlingshirn24,Brinkman73,Capizzi84,Kira98,
Huber2001,Loevenich02,Kaindl2003,Almand-Hunter2014,Stolz2021}.

Naturally matching the characteristic picosecond time scale, an intense terahertz electromagnetic pulse (1~THz = 1~ps$^{-1}$) is used in this work to strongly interact with the nonequilibrium many-body states in order to reveal their nonlinear characteristics, while linear THz probes have been widely applied previously (see e.g. \cite{Huber2001,Kaindl2003,Suzuki09}).
We observe strong THz third-harmonic emission of the nonthermal states after the initial optical creation of electron-hole pairs, and utilize the nonlinear response to monitor the time-dependent evolution of the far-from-equilibrium many-body states.
By carrying out a systematic investigation of the observed THz third-harmonic generation as a function of pump-drive delay, optical pump fluence, and THz field strength, we find distinct, and a rich set of THz nonlinear characteristics of the strongly-correlated electron-hole many-body states. 

Cuprous oxide Cu$_2$O is a highly suitable semiconductor system chosen for the study of the involved many-body dynamics. The canonical Rydberg excitons with principal quantum number up to $n=30$ were observed in Cu$_2$O \cite{Nature2014giant,PRB2021giant,Nat.Mater.2022rydberg}, while typically up to $n = 5$ in two-dimensional semiconductor crystals \cite{Chernikov14}.
The lowest-lying exciton state is sufficiently long-lived, i.e. $\sim 0.1-10$~$\mu$s \cite{Mysyrowicz79,Yoshioka10}, for a possible realization of excitonic Bose-Einstein condensation \cite{PRL1990evidence,RPP2014bose, Nat.Commu.2022observation}.
Whereas bright excitons can be identified in conventional linear-response spectroscopy of thermal equilibrium state by measuring near-infrared/visible absorption \cite{Nature2014giant} or photoluminescence \cite{PRB2021giant},
an access to dark excitonic states requires nonlinear and/or nonequilibrium state spectroscopy that employs two-photon processes \cite{PRB2018High-resolution,PRL2020Rydberg,PRB2020Two-photon} or intraexciton optical transitions \cite{PRL2005study,PRL2006stimulated,PRL2008terahertz}. 
In particular, a resonant optical transition excitation of the $1s$ to $2p$ exciton states was resolved, at 10~ps after the creation of unbound electron-hole pairs, which determined a characteristic time scale of the $1s$ Rydberg exciton formation \cite{PRL2005study,PRL2008terahertz}.
  
\begin{figure}[t]
\centering 
\includegraphics[width=0.9\linewidth]{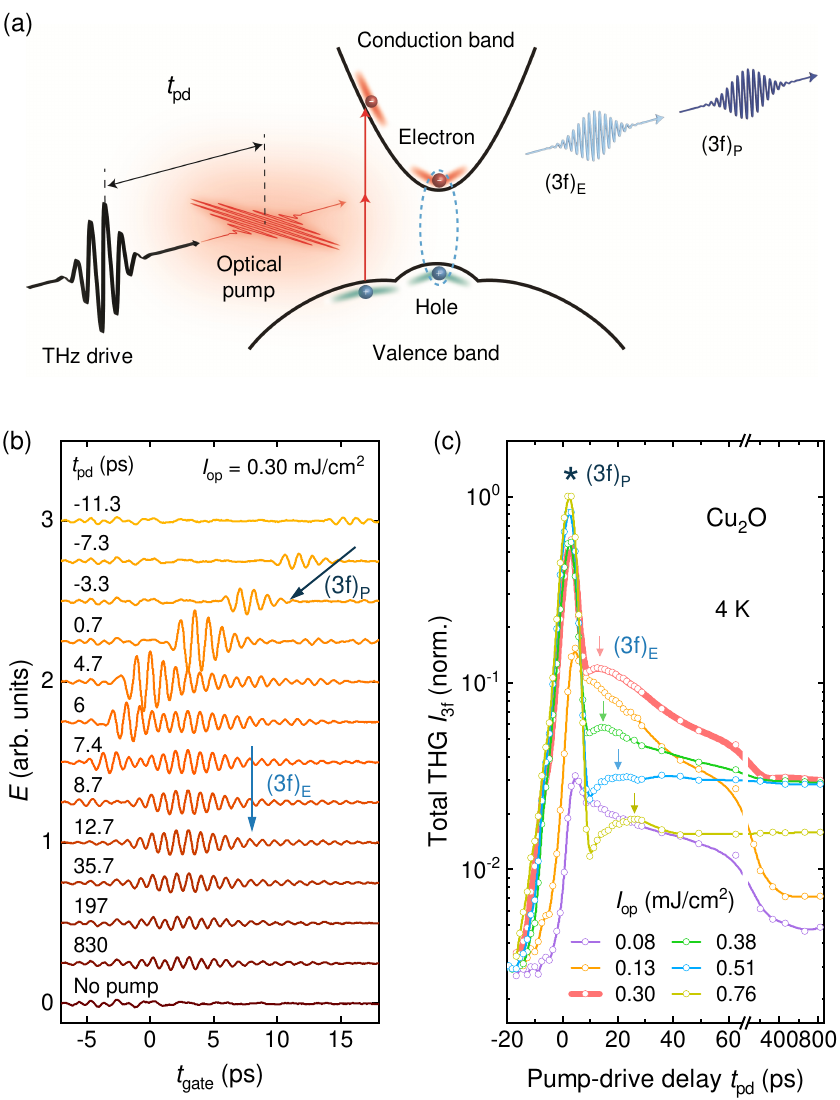}
\caption{ 
(a) Illustration of optical-pump THz-drive third-harmonic-generation measurement.
(b) Time-domain traces of emitted THz electric field $E$ for various optical-pump THz-drive delays $t_\text{pd}$ at an optical-pump fluence of $I_\text{op} = 0.30$~mJ/cm$^2$ and an $f=0.35$~THz drive peak field of $E_{f} = 120$~kV/cm. 
The bottom trace represents the background without optical pump.
The emitted field after a $3f$ bandpass filter was gate-detected by electro-optic sampling with an 800~nm laser pulse for tunable gate time $t_\text{gate}$.
(c) Total THG intensity $ I_\text{3f} $ versus $ t_\text{pd} $ for various pump fluences.
}
\label{Fig1} 
\end{figure}

We use a 100~fs laser pulse (800~nm, 1.55 eV) to induce two-photon excitation of unbound electron-hole pairs in a naturally grown single crystal Cu$_2$O whose band gap is 2.172~eV \cite{PRL2008terahertz}.
Generated by optical rectification of 800~nm pulse from the same laser system in a LiNbO$_3$ crystal \cite{Hebling_02}, an intense THz drive pulse is naturally synchronized to the pump pulse with a tunable pump-drive time delay $t_\text{pd}$ [see Fig.~\ref{Fig1}(a)].
Due to the centrosymmetric cubic structure of Cu$_2$O, the intraband even-order nonlinear susceptibilities vanish, hence we focus on the third-order THz nonlinear responses.
To prepare a narrowband THz pulse with a central frequency $f$ and to detect third-order harmonic generation at $3f$, we use corresponding THz bandpass filters with a 20\% bandwidth.
The time traces of the THz pulses are detected via electro-optic sampling in a ZnTe crystal \cite{wu1995free}.
The sample temperature is 4~K in a helium flow optical cryostat.

\textit{Results} - Strong terahertz third harmonic generation (THG) is observed by driving Cu$_2$O with an intense THz pulse after an 800~nm optical excitation for varying the optical-pump THz-drive delay $t_\text{pd}$.
Figure~\ref{Fig1}(b) presents the time-domain traces of the emitted THG electric field at various $t_\text{pd}$'s for an optical-pump fluence of $I_\text{op} = 0.30$~mJ/cm$^2$ and a $f=0.35$~THz drive with a peak field of $E_{f} = 120$~kV/cm.
Whereas without optical excitation no THG is observed, a most efficient third-harmonic yield occurs at $t_\text{pd}=3.4$~ps [Fig.~\ref{Fig1}(c)].
The negative values of $t_\text{pd}$ in the representation correspond to the experimental settings that the optical pump pulse reaches the sample after the peak of the THz drive field.
Even at $t_\text{pd}=-11.3$~ps a weak THG trace labelled as $(3f)_P$ is still discernible i.e. after $t_\text{gate}=13$~ps [Fig.~\ref{Fig1}(b)], because the optical excitation pulse catches up the tail of the multicycle THz drive pulse whose duration is about 20~ps.
With increasing $t_\text{pd}$, the THG transient $(3f)_P$ shifts continuously towards smaller values of $t_\text{gate}$, as indicated by the dark blue arrow in Fig.~\ref{Fig1}(b).

A distinctly different THG feature occurs at positive $t_\text{pd}$.
At $t_\text{pd}=4.7$~ps the duration of the THG trace appears to be longer than that at $t_\text{pd}=0.7$~ps, which becomes even longer at $t_\text{pd}=6$~ps.
This is not due to stretch of the THG transient in the time domain, but reflects the occurrence of a second THG transient after $(3f)_P$.
At a longer pump-drive delay of $t_\text{pd}=7.4$~ps the second THG transient, as labelled by $(3f)_E$, is temporally well-separated from the $(3f)_P$ transient, hence can be directly identified in our time-resolved measurement.
The $(3f)_E$ transient does not shift with the optical pump pulse in the time domain, but appears only when the THz drive pulse arrives, even for a quite long pump-drive delay, see e.g. $t_\text{pd}=830$~ps in Fig.~\ref{Fig1}(b).
Such a long delay (i.e. $\sim 1$~nanosecond) corresponds to the characteristic time scale of the long-lived $1s$ Rydberg exciton state in Cu$_2$O \citep{PRL2008terahertz}.
As will be further discussed below, we can assign the $(3f)_E$ transients as the featuring third-harmonic responses of excitons under the THz drive, while the $(3f)_P$ radiation as THz THG of electron-hole plasma.

A second distinct behavior of the excitonic THG transient $(3f)_E$, which is characteristic for the nonequilibrium many-body states, is the nonmonotonic dependence of its intensity on the pump-drive delay $t_\text{pd}$.
As can be directly seen from the time-domain signal [Fig.~\ref{Fig1}(b)], the $(3f)_E$ intensity at $t_\text{pd}=12.7$~ps is not only stronger than those at longer pump-drive delays (e.g. $t_\text{pd}=35.7$~ps), but also than that of $t_\text{pd}=8.7$~ps.
This is more evident in the representation of the total integrated THG intensity $I_{3f}$ in Fig.~\ref{Fig1}(c).
At a critical value of $I^c_\text{op}=0.30$~mJ/cm$^2$ the maximum THG intensity achieved at $t_\text{pd}=3.4$~ps, as marked by the asterisk in Fig.~\ref{Fig1}(c), is followed by a second smaller maximum at $t_\text{pd}=12.7$~ps, as indicated by the arrow.
Hence the second maximum as being due to the excitonic THG reflects the typical time scale of exciton formation \cite{PRL2008terahertz} after the excitation of electron-hole pairs, while the first maximum corresponds to the THG of the optically excited electron-hole plasma under the THz drive.

A third nonlinear characteristic of the Coulomb correlated nonequilibrium many-body states is featured by the nonmonotonic  dependence of the excitonic THG signal $(3f)_E$ on the electron-hole density, which is tunable by varying the pump-pulse fluence.
As shown in Fig.~\ref{Fig1}(c), with increasing pump fluence, the maximum third-harmonic yield corresponding to the electron-hole plasma $(3f)_P$ increases monotonically, while the position of this maximum shifts slightly to shorter pump-drive delays.
In contrast, the maximum of the excitonic THG $(3f)_E$ reaches a largest value at $I^c_\text{op}=0.30$~mJ/cm$^2$, and the maximum position shifts monotonically towards longer pump-drive delays (i.e. from 12.7~ps at 0.30~mJ/cm$^2$ to 26.0~ps at 0.76~mJ/cm$^2$, see arrows in Fig.~\ref{Fig1}(c)).
For $I_\text{op}<I^c_\text{op}$ the $(3f)_E$ maximum merges into the $(3f)_P$ peak thus cannot be resolved. 

\begin{figure}[t]
\centering 
\includegraphics[width=1\linewidth]{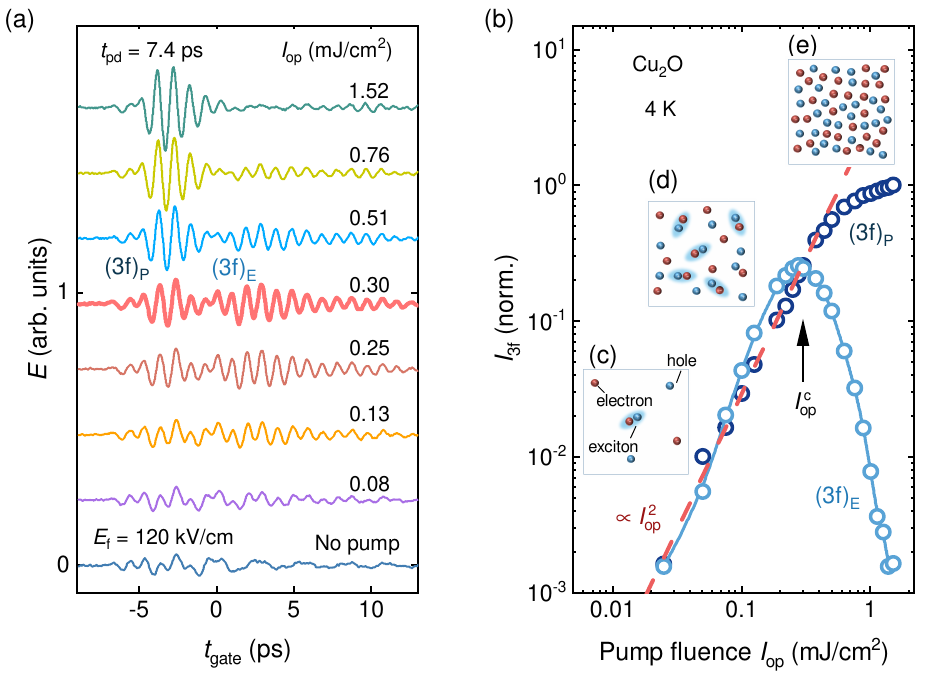}
\caption{ 
(a) Emitted THz field traces at various optical-pump fluences for $t_\text{pd} = 7.4$~ps and a $0.35$~THz drive peak field of $120$~kV/cm.
Two temporally separate THG transients are resolved and marked by $(3f)_P$ and $(3f)_E$, respectively.
(b) Integrated intensity of these two transients versus 
pump fluence. 
The dashed line indicates quadratic dependence $I_\text{3$ f $} \propto  I^2_\text{op}$ for $I_\text{op}<I^c_\text{op}$.
(c) Sketch of unbound charges and exciton in the dilute density limit.
(d) Illustration of strongly correlated many-body states of electron-hole plasma and excitons with higher charge density.
(e) In the high-density limit, the Coulomb interaction within an exciton is completely screened, leaving behind a strongly interacting electron-hole plasma.}
\label{Fig2} 
\end{figure}

To further reveal nonlinear THz excitonic sensing of the many-body states, we fix the pump-drive delay at $t_\text{pd}=7.4$~ps, where the $(3f)_P$ and $(3f)_E$ transients are well discriminated in the time domain, and study their evolution with varying pump fluence (or equivalently the charge density).
As shown in Fig.~\ref{Fig2}(a), the time-domain THG signal of $(3f)_P$ increases monotonically with the pump fluence, whereas the $(3f)_E$ signal first increases until the critical value of $I^c_{op}=0.30$~mJ/cm$^2$ and then decreases.
The pump-fluence dependence of the integrated intensity $I_{3f}$ of these two temporally separated THG transients $(3f)_P$ and $(3f)_E$ is presented in Fig.~\ref{Fig2}(b).
At the lowest fluences the THG intensity of the two transients is comparable and increases monotonically, which reflects a concomitant increase of the optically excited charge-carrier and exciton densities.
The dependence on the optical pump fluence follows essentially a quadratic behavior, i.e. $I_{(3f)_P}\propto I_\text{op}^2$ [see dashed line in Fig.~\ref{Fig2}(b)], which is consistent with the two-photon excitation of electron-hole pairs across the band gap [Fig.~\ref{Fig1}(a)] \cite{PRL2008terahertz}.
At high fluences, i.e. $I_\text{op}>I^c_\text{op}$, the decrease of the excitonic THG intensity $(3f)_E$ in contrast to the continuous increase of plasma third-harmonic yield $(3f)_P$ is a nonlinearity signature for a Mott phase of excitons due to the high density.
Also the increase of the plasma THG deviates from a perturbative quadratic dependence, reflecting an enhanced Coulomb interaction among the electrons and holes at high densities.

Figure~\ref{Fig2}(c)-(e) illustrate the evolution of the many-body states excited by the 800~nm laser pulses as a function of charge-carrier density (or equivalently, of laser fluence). 
At very low pump fluences [Fig.~\ref{Fig2}(c)], both the electron-hole plasma density and the exciton density increase continuously, which, under the THz drive, leads to a quadratic increase of the third-harmonic radiation [see Fig.~\ref{Fig2}(b), dashed line].
The excitons in the presence of the charged particles (i.e. electrons and holes) are not independent from each other, but constitute together a highly correlated many-body state due to the strong Coulomb interaction [Fig.~\ref{Fig2}(d)]. 
The interaction between electrons and holes in the plasma can lead to the formation of excitons, while at the same time excitons are scattered into electrons and holes in the plasma.
Moreover, the electron-hole interaction within an exciton is screened by the other charges in the plasma and excitons, hence the excitons become more and more fragile with increasing charge density.
Above a critical density (i.e. corresponding to $I^c_\text{op}=0.30$~mJ/cm$^2$ in Fig.~\ref{Fig2}(b)), the excitons start to decay, which corresponds to the decrease of $I_{(3f)_E}$, and finally evolve into a Mott phase where only plasma remains [Fig.~\ref{Fig2}(e)].  
For these strongly correlated many-body states, which are formed just few picoseconds after the optical excitation, our time-resolved measurement of THz nonlinear responses can clearly unveil the signature of the excitons in presence of the plasma, thereby revealing their evolution towards the Mott phase.

\begin{figure}[t]
\centering 
\includegraphics[width=0.9\linewidth]{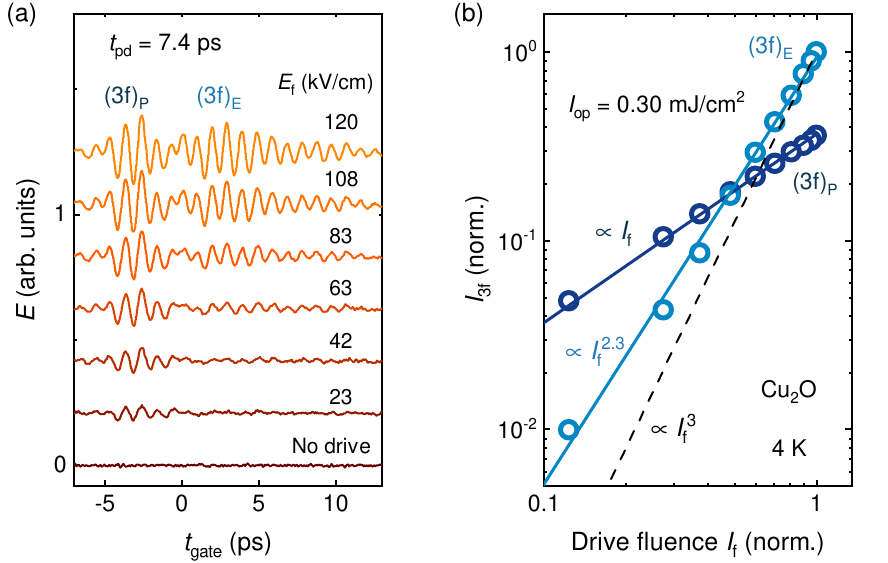}
\caption{
(a) Emitted THz field $E$ measured with different drive fields for $t_\text{pd} = 7.4$~ps and $I_{op}=0.30$~mJ/cm$^2$.
(b) The obtained third-harmonic yield of $(3f)_P$ and $(3f)_E$ exhibits very different dependencies on the drive fluence, i.e. $I_{(3f)_E} \propto I^{2.3}_f$ whereas $I_{(3f)_P} \propto I_f$.
The dashed line indicates a perturbative power-law dependence $I_\text{3f} \propto I_\text{f} ^{3} $.  
} 
\label{Fig3} 
\end{figure}

The distinction of the nonlinear THz excitonic dynamics is not only manifested by the charge-density dependent evolution, but also by a characteristic difference of its dependence on the THz drive field.
Figure~\ref{Fig3}(a) presents the emitted third-harmonic radiation at various THz drive field strengths for a fixed optical pump fluence of the critical value $I^c_\text{op}=0.30$~mJ/cm$^2$ and a pump-drive delay of $t_\text{pd} = 7.4$~ps, where the excitonic THG is clearly distinguished from the plasma THG in the time domain.
Without a THz drive no $3f$ radiation is detectable within our experimental uncertainties.
Hence both the THz drive and the optical pump [see also Fig.~\ref{Fig1}(b)] are required for the observation of the THz third-harmonic radiation.

The plasma THG $I_{(3f)_P}$ starts to appear already at a relatively low THz drive field (e.g. 23~kV/cm), while the excitonic THG $I_{(3f)_E}$ is hardly discernible. 
With increasing drive field strength, both the plasma and the exciton THG increases continuously, in contrast to the nonmonotonic dependence on the optical pump fluence (see Fig.~\ref{Fig2}).
In comparison with that of plasma, the enhancement of the excitonic THG is much faster, which becomes even dominant over the plasma THG at the highest drive fields available to our experiment.
The temporal separation of the exciton and plasma THG enables a separate evaluation of their intensity, the result of which is presented as a function of the THz drive pulse fluence $I_f$ in Fig.~\ref{Fig3}(b).
Interestingly, neither of the two THG transients exhibits a cubic power law dependence on $I_f$, which is in clear contrast to a perturbative third-order nonlinear response. 
Whereas the plasma THG exhibits essentially a linear dependence $I_{(3f)_P} \propto I_f$, the dependence of the excitonic THG follows a nonperturbative power-law dependence
$I_{(3f)_E} \propto I^{2.3}_f$, which refute a trivial interpretation by a perturbative nonlinear susceptibility.

\textit{Discussions} - Our results have provided a systematic THz nonlinear characterization of the strongly correlated nonequilibrium electron-hole many-body states. 
The observed distinct difference in the THz fluence dependencies of the plasma and the exciton THG indicates two different mechanisms for the generation of the third harmonic radiation. As will be discussed below, one mechanism roots in the nonparabolicity of the exciton dispersion relation, while the other is because of the highly nonthermal nature of the far-from-equilibrium electron-hole plasma. 

In Cu$_2$O, different from the lowermost conduction band $ \Gamma_{6}^{+} $ with a parabolic dispersion relation in the vicinity of the $\Gamma$ point, the dispersion relation of the uppermost valence band $ \Gamma_{7}^{+} $ is characterized by a clear nonparabolicity due to spin-orbit coupling [see Fig.~\ref{Fig1}(a) for an illustration] \cite{JPCM2008electronic,PRB2016deviations}.
In the Hamiltonian in addition to the parabolic terms $H^\text{P}(k_x,k_y,k_z) = -a_1k^2_x-a_2k^2_y-a_3k^2_z$ with coefficients $a_{1,2,3}$ and the quasi-momentum components $k_{x,y,z}$, the nonparabolic terms up to the second order are in general of the form
$H^\text{NP}(k_x,k_y,k_z)=-c_1|k_x||k_y|-c_2|k_y||k_z|-c_3|k_z||k_x|$
where the coefficients $c_{1,2,3}$ arise due to spin-orbit and orbit-orbit coupling \cite{PRB2016deviations}.
To clarify the responsibility of the nonparabolicity for the observed THG, we can consider a simplified model for a linearly polarized THz drive field along the $x$-axis with an idealized sinusoidal waveform $E_f\sin(2\pi f t)$.
Under the drive of this field, the parabolic terms will lead to a velocity of the charged quasiparticles 
$v^P(t) \propto \frac{\partial H^P}{\partial k_x} = -2a_1k_x \propto \cos(2\pi ft)$, hence the emitted THz electric field
$\propto \frac{\partial j^P(t)}{\partial t} \propto \sin(2\pi f t)$ with the time-dependent current density $j^P(t) = nv^P(t)$, which is again a sinusoidal waveform with the fundamental frequency $f$.
In contrast, the nonparabolic terms give rise to
$v^{NP}(t) \propto \frac{\partial H^{NP}}{\partial k_x} = \mp (c_1|k_y| + c_3|k_z|)$, which for the linearly polarized THz field results in a time-dependent current density $j^{NP}(t)=nv^{NP}(t)$ of a rectangular waveform in an idealized scenario. Consequently, the emitted THz electric field, which is proportional to $\frac{\partial j^{NP}(t)}{\partial t}$, contains high-order harmonics.
This idealized model has already captured the essence of the physics, although the energy loss due to inelastic scattering in real materials is neglected. 
Within the framework of Boltzmann transport theory, one can take into account scattering processes and show that the nonperturbative power-law dependence results from a strong THz field driven nonlinear kinetics (see e.g. Refs.~\cite{NC2020Non,Dantas2021}).  
Based on this mechanism, we can naturally ascribe the observed THz THG to the field-driven kinetics of holes from the nonparabolic band $ \Gamma_{7}^{+} $, both for excitons and plasma.
It is worth noting that a fully quantum-mechanical description of the many-body nonlinear kinetics driven by intense THz field remains an outstanding theory problem \cite{Borsch2023}.

\begin{figure}[t]
\centering 
\includegraphics[width=1\linewidth]{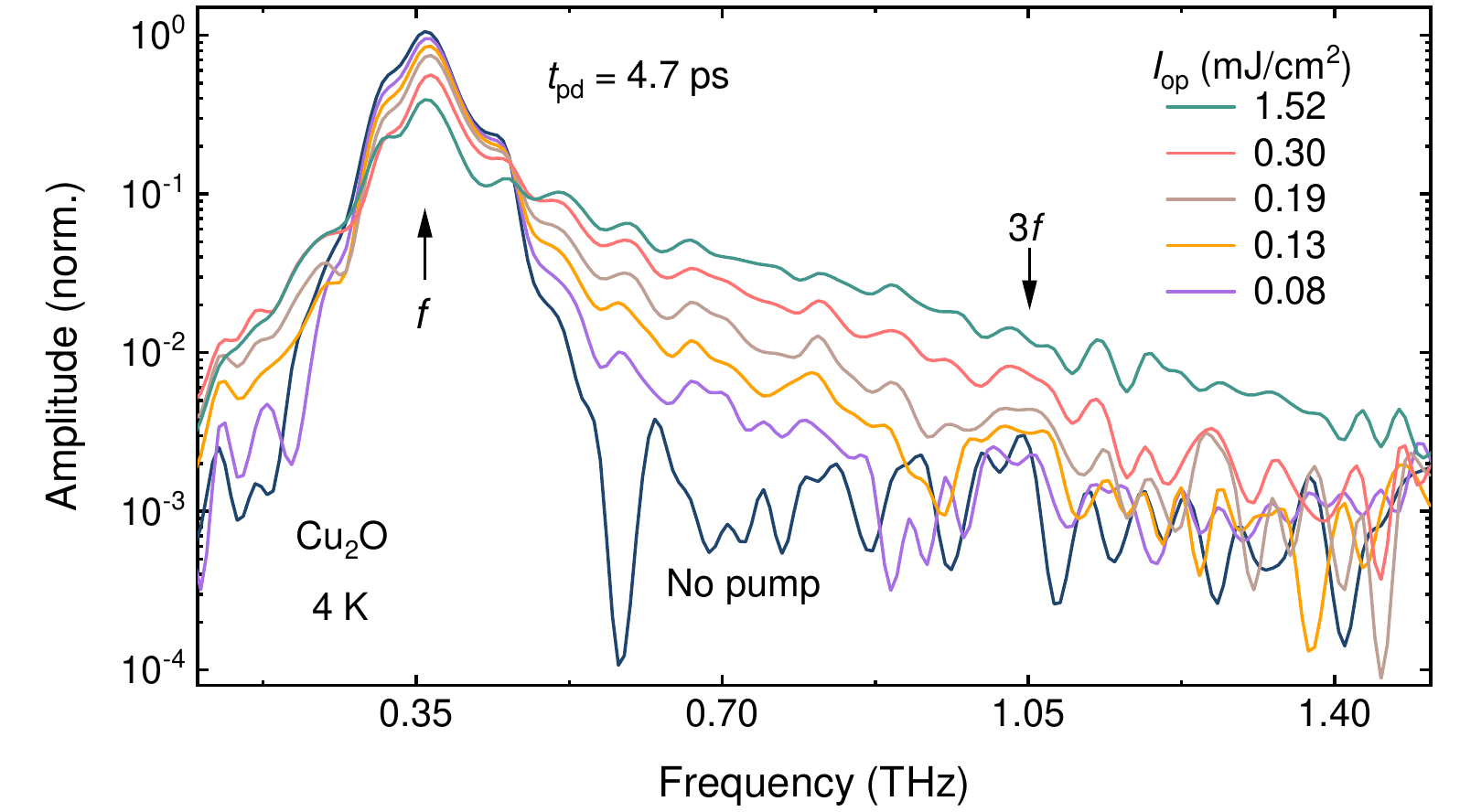}
\caption{
Broadband THz emission spectrum of optically excited electron-hole plasma under an $f=0.35$~THz drive at a fixed, short pump-drive delay of $t_\text{pd}=4.7$~ps for various pump fluences $I_{op}$.
The spectrum without optical pump characterizes the spectral profile of the transmitted THz drive pulse, where above $\sim 0.5$~THz reflects the experimental noise level.
With increasing fluence the emission spectra are broadening, which effectively enhances the spectral weight at $3f$.  
} 
\label{Fig4} 
\end{figure}

The observed linear dependence of the plasma THG intensity (i.e. $I_{(3f)_P} \propto I_f$) deviates rather far from the perturbative cubic power law, strongly suggesting a different mechanism than for the excitonic THG (i.e. $I_{(3f)_E} \propto I^{2.3}_f$).
To investigate the plasma THG further, we shift the pump-drive time delay to $t_\text{pd}=4.7$~ps, where the emitted THG is dominated by the plasma contribution [see Fig.~\ref{Fig1}(b)], and measure the emitted THz field in the time domain without a \textit{3f} bandpass filter for various pump fluences.
The derived THz emission spectra in the frequency domain are presented in Fig.~\ref{Fig4}.

Without the optical pump, we observe just the transmission of the THz drive pulse through the sample [see also Fig.~\ref{Fig1}(b)], which is centered at the frequency of $f=0.35$~THz without resolvable spectral weight at $3f$ above the noise level.
In contrast, with an optical pump, already at a low pump fluence of 0.08~mJ/cm$^2$, one can see an evident change of the THz emission spectrum, which is characterized by a broadening beyond the spectrum of the THz drive pulse, both towards the lower and higher frequencies.
With increasing pump fluence, the broadening is further enhanced in the frequency domain.
At the highest pump fluence, the high-frequency limit of the broadband emission spectrum reaches up to about 1.5~THz above the experimental noise level.
These results show that the stimulated THz emission due to the plasma is of broadband nature.
This can be understood based on the experimental fact that the electron-hole plasma is photoexcited rather instantaneously by the 100~fs optical pump pulse, in comparison with the few-picosecond long THz drive pulse.
Hence, in the time domain the THz drive pulse experiences a sudden absorption by the electron-hole plasma as soon as the optical pulse arrives at the sample, which in the frequency domain corresponds to a broadband THz emission.
This scenario is confirmed by the observation, as shown in Fig.~\ref{Fig4}, that with increasing optical pump fluence the peak intensity at the THz drive frequency of 0.35~THz drops concomitantly with the broadening of the emission spectrum.

To conclude, we observed strong and characteristic terahertz nonlinear responses for optically excited, transient and highly correlated electron-hole many-body states in a well-known semiconductor Cu$_{2}$O.
Efficient and nonpertubative terahertz third-harmonic generation has been observed few picoseconds after the optical creation of electron-hole pairs, which evidences the formation of bound electron-hole pairs - excitons - despite the presence of Coulomb collisions with unbound electrons and holes.
Nonetheless, when the electron-hole density is sufficiently high, excitons are hardly formed from the strongly interacting many-body states, as manifested by their vanishing terahertz third-harmonic yield.
Our findings provide an efficient approach to sense electron-hole many-body states by utilizing terahertz nonlinear excitonic responses, and pave the way for the understanding of quantum many-body effects in various material classes \cite{Wilson2021,Goulielmakis2022,Borsch2023,Valmispild2024,Heide2024}.

\begin{acknowledgements}
We acknowledge support by the European Research Council (ERC) under the Horizon 2020 research and innovation programme, Grant Agreement No. 950560 (DynaQuanta), and by the German Research Foundation (DFG) through grant No. 504522424.
\end{acknowledgements}

\bibliographystyle{apsrev4-2}
\bibliography{Cu2O_Paper_bib}

\end{document}